\documentstyle[11pt]{article}

\newcommand{\be}{\begin{equation}}
\newcommand{\ee}{\end{equation}}
\newcommand{\ba}{\begin{array}}
\newcommand{\ea}{\end{array}}
\newcommand{\bc}{\begin{center}}
\newcommand{\ec}{\end{center}}
\newcommand{\bi}{\begin{itemize}}
\newcommand{\ei}{\end{itemize}}

\newcommand{\disregard}[1]{{}}

\def\bild#1\over#2{\mathrel{\mathop{\kern0pt #1}\limits_{#2}}}

\begin{document}
{\bf  TOPOLOGICAL 2-DIMENSIONAL QUANTUM MECHANICS\rm}
\vskip 2cm
{\bf Alain DASNI\`ERES de VEIGY and St\a'ephane OUVRY \rm }\footnote{\it  and
LPTPE, Tour 16, Universit\'e Paris  6 / electronic e-mail: OUVRY@FRCPN11}
\vskip 1cm
{Division de Physique Th\'eorique \footnote{\it Unit\a'e de Recherche  des
Universit\a'es Paris 11 et Paris 6 associ\a'ee au CNRS},  IPN,
  Orsay Fr-91406}
\vskip 3cm
Abstract: We define a Chern- Simons Lagrangian for a system of
planar particles topologically
interacting at a distance.  The anyon model appears as a particular case
where all the particles are identical. We propose exact N-body eigenstates,
set up a perturbative algorithm, discuss the case where some particles
are fixed on a lattice, and also consider curved manifolds.

PACS numbers: 05.30.-d, 11.10.-z
\vskip 3cm

IPNO/TH 92-101
December 1992

\vfill\eject

1. INTRODUCTION

The anyon [1] (quantum mechanical, non relativistic) model is a
fascinating system to study.

Firstly, it is a Aharonov-Bohm [2] system with no classical counterpart,
in which planar particles interact at a distance topologically. Secondly,
the particles being identical, the effect of the interaction
is to make their statistics intermediate [1,3], nor Bose, neither Fermi.

Little is known about anyons despite huge efforts devoted
to their study.
In the N-anyon case with harmonic attraction to the origin,
linear eigenstates have been constructed [4], but they are
known to be only part of the spectrum. Perturbative approaches [5] have
undercovered
a very complex structure for the equation of state. Finally, several
numerical analysis have been developped [6].

Here, one is going to drop the indistinguishibility of the particles
and propose a general model of particles topogically interacting at
a distance \`a la  Aharonov-Bohm.

2. THE FORMALISM

Consider the density of Lagrangian for N particles moving in plane
minimally coupled to vector gauge fields $ A_{\alpha}^{\mu}(\vec r)$
\be\label{1} L_N = \sum_{i=1}^N \big({1\over 2}
{m{\vec v_i}^2}+ \sum_{\alpha=1}^Ne_{\alpha i}(\vec A_{\alpha}(\vec r_i)
\vec v_i - A_{\alpha}^o(\vec r_i))\big) + \sum_{\alpha,\beta=1}^N
{\kappa_{\alpha\beta}\over 2}\epsilon_{\mu\nu\rho}\int
A_{\alpha}^{\mu}\partial^{\nu}A_{\beta}^{\rho} d\vec r\ee

In (\ref{1}), the index $i$ refers to the number affected to a given particle
among $N$, the indices $\mu,\nu,\rho,\ldots $ correspond to
the 3-dimensional Lorentz degrees freedom ($\mu=0$ denotes the time direction,
$\mu=1 ,2$ the space directions), and $\alpha,
\beta,\ldots $ label some internal degrees of freedom carried by
the vector fields ($1<\alpha<N$).
The $e_{\alpha i}$'s are the electromagnetic couplings (topological charges)
between the matter particles $i$ and the gauge fields $\alpha$
at position $\vec r_i$. The $\kappa_{\alpha\beta}$ are the
Chern-Simons\footnote{Here
we concentrate on abelian Chern-Simons gauge fields. The generalization
to the non-abelian case could be studied along the same lines.} self-couplings
of the gauge fields $\alpha,\beta$.

One can easily see that under the  gauge transformation
\be\label{2} \delta A^{\mu}_{\alpha}=\partial^{\mu}\Lambda_{\alpha}\ee
the density of Lagrangian (\ref{1})
changes by a time derivative, namely
$\delta \epsilon_{\mu\nu\rho}
A_{\alpha}^{\mu}\partial^{\nu}A_{\beta}^{\rho} =
\partial^{\mu}(\epsilon_{\mu\nu\rho}\Lambda_{\alpha}\partial^{\nu}
A^{\rho}_{\beta})$.
Also, by the very definition (\ref{1}) one can restrict
to $\kappa_{\alpha\beta}=\kappa_{\beta\alpha}$, i.e to a
symmetric matrix $[\kappa]$.

One  proceeds by eliminating the time components of the gauge fields
$A_{\beta}^o$
by varying the Lagrangian with respect to them
($\delta L_N/\delta A_{\beta}^o=0$). One gets
\be\label{3}\sum_{i=1}^N e_{\beta i}\delta^2(\vec r-\vec r_i) =
\sum_{\alpha=1}^N\kappa_{\alpha\beta}\epsilon_{o\mu\nu} \partial^{\mu}
 A_{\alpha}^{\nu}\ee
where explicit use of the symmetry of $[\kappa]$ has been made.

The magnetic field $B_{\alpha}=\epsilon_{o\mu\nu} \partial^{\mu}
 A_{\alpha}^{\nu}$ appears in the right hand side of (\ref{3}). Using the
 bidimensional identity
$\vec \partial .{\displaystyle \vec r\over r^2}=2\pi \delta^2(\vec r)$ one
finds
in the Coulomb gauge $\vec \partial . \vec A_{\alpha}(\vec r)=0$
\be\label{4}
\sum_{\alpha}\kappa_{\alpha\beta}\vec A_{\alpha}(\vec r)=\sum_i{e_{\beta i}
\over 2\pi}\vec k
\times {\vec r-\vec r_i\over(\vec r-\vec r_i)^2}\ee
($\vec k$ is the unit vector
perpendicular to the plane).
This equation can be symbollically rewritten as
$[\kappa][\vec A_{\alpha}(\vec r)]=[{e/ 2\pi}]
[\vec k\times {(\vec r-\vec r_i)/(\vec r-\vec r_i)^2}]$ where
$[\vec A_{\alpha}(\vec r)]$ and $[\vec k\times
{(\vec r-\vec r_i)/(\vec r-\vec r_i)^2}]$ are one column vectors on which
the matrices $[\kappa]$ and $[e]$ act. If one assumes that
$[\kappa]$ is inversible one gets
\be\label{5} [\vec A_{\alpha}(\vec r)]=
[\kappa]^{-1}[{e\over 2\pi}][\vec k\times {\vec r-\vec r_i\over(\vec r-\vec
r_i)^2}]\ee

The Hamiltonian corresponding to $L_N$ stems from
\be\label{6} \vec p_i\equiv {\partial L_N\over\partial\vec v_i}=
m\vec v_i+\sum_{\alpha}e_{\alpha i}\vec A_{\alpha}(\vec r_i)\ee
One gets
\be \label{7} H_N={1\over 2m}\sum_i{(\vec p_i-
\sum_{\alpha}e_{\alpha i}\vec A_{\alpha}(\vec r_i))^2}\ee
If one defines ${\cal \vec A}_i(\vec r)\equiv
\sum_{\alpha}e_{\alpha i}\vec A_{\alpha}(\vec r)$
one sees that $[{\cal \vec A}_i(\vec r)]
=[e]^t[\kappa]^{-1}[{e/ 2\pi}][\vec k\times {(\vec r-\vec r_i)/(\vec r-\vec
r_i)^2}]$
at position $\vec r_i$ enters the definition of $H_N$.

Let us consider the coupling matrix $[\alpha]\equiv
[e]^t[\kappa]^{-1}[{e/ 2\pi}]$. It is by definition a symmetric matrix.
The usual anyon model is nothing
but taking $[\kappa]$ and $[e]$ to be one-dimensional matrices (i.e.
one single gauge field)
with a single anyonic
coupling constant $\alpha ={\displaystyle e^2\over \displaystyle 2\pi\kappa}$
between the "flux" $\phi={\displaystyle e\over \displaystyle \kappa}$ and the
"charge"
$e$ carried
by each anyon. Singular self-interaction, which are present
in $\vec A(\vec  r)$ at position $\vec r=\vec r_i$, have to be left aside.

Here, we get a general model where the couplings $\alpha_{ij}
=\alpha_{ji}$
can depend on $i$ and $j$.
One notes that
${\cal \vec A}_i(\vec r)$ is correctly defined at $\vec r=\vec r_i$
if and only if one asks for the matrix $[\alpha]$ to
have its diagonal elements equal to $0$.
Thus one ends up with $N(N-1)/2$ independant anyonic
coupling constant (the entries of $[\alpha]$) and a Hamiltonian
\be\label{8}  H_N= {1\over 2m}\sum _{i=1}^{N}(\vec{p}_i -
{\cal \vec A}_i)^2\ee
where the gauge field  ${\cal \vec A}_i= \sum_{j\ne i}\alpha_{ij}
{\displaystyle \vec k \times \vec r_{ij}\over \displaystyle r_{ij}^2}$
with $\vec r_{ij}=
\vec r_i - \vec r_j$.

In this more general point of view, the anyon model can be  recovered
by taking all the $\alpha_{ij}$ equal to $\alpha$. But now, one gets
 as a bonus that
${\cal \vec A}_i(\vec r)|_{\vec r=\vec r_i}$ is defined at the position
of the particles, which was not the case
in the original formulation.
A quantum mechanical model of flux tubes $\phi_j$
interacting \`a la Aharonov-Bohm with electric charges $e_i$ would correspond
to taking $e_{\alpha i}=e_i$ implying
$\phi_j=(\sum_{\alpha\beta}
[\kappa]^{-1}_{\alpha\beta})e_j$ and $\alpha_{ij}=
{\displaystyle 1\over \displaystyle 2\pi}
e_i(\sum_{\alpha\beta}
[\kappa]^{-1}_{\alpha\beta})e_j$.

$[\kappa]$ matrices have been already used [7]
in order to reproduce fractionnal values for the Hall conductivity given as
$\sigma_H=\sum_{\alpha\beta}
[\kappa]^{-1}_{\alpha\beta}$.
In this sghlightly different
context (in particular there is no kinetic term for
matter), one insists on a matrix $[\kappa]$ with integer entries
in order to reproduce the quantum numbers (statistics, charge) of the electron.

Here the entries are not constrained to be integers, in $[\kappa]$ as well
as in $[e]$.

As in the anyon model,
${\cal \vec A}_i(\vec r_i)$ is pure gauge,  the singular gauge parameter being
\be\label{9}{\cal \vec A}_i(\vec r_i)=\vec \nabla_i(\sum_{k<l}
\alpha_{kl}\theta_{kl})\ee
where $\theta_{kl}$ is the relative angle of the  particles $k$ and $l$.
The strong analogy between the anyon model and the model proposed above will
allow for the generalization of interesting results of the former to
the latter.

3. SOME EXACT AND PERTURBATIVE RESULTS

i) Exact results

The structure of the Hamiltonian $H_N(\alpha_{ij})$ given in
(\ref{8}) allows for the
following general comment. Suppose one has an eigenstate $\psi(\alpha_{ij})$
of energy $E(\alpha_{ij})$. By the virtue of the gauge transformation
\be\label{10} \psi'(\alpha_{ij}) =\exp(i\sum_{k<l} m_{kl}\theta_{kl})
\psi(\alpha_{ij})\ee
one finds that $\psi'(\alpha_{ij})$ is a monovalued
eigenstate of the Hamiltonian
$H_N(\alpha_{ij}-m_{ij})$ with the same energy $E(\alpha_{ij})$.
This implies that $\psi'(\alpha_{ij}+m_{ij})$ is a monovalued
eigenstate of the original
Hamiltonian $H_N(\alpha_{ij})$ with energy $E(\alpha_{ij}+m_{ij})$.
Thus as soon one knows an exact eigenstate, one can associate to it
 a tower of orbital eigenstates indexed by the quantum numbers $m_{ij}$.

Another consequence is that if one considers the $m_{ij}$'s as
gauge parameter coefficients (not as actual quantum numbers), and choose
$m_{ij}=E[\alpha_{ij}]$,  where $E[\alpha_{ij}]$ is the integer part of
$\alpha_{ij}$, the gauge transformed Hamiltonian
$H_N(\alpha_{ij} - m_{ij} )$ is then defined in terms of the
couplings $\alpha_{ij}'=\alpha_{ij}-E[\alpha_{ij}]$ with the constraint that
$\alpha_{ij}'\in [0,1]$. Clearly,
the physics described by both models is the same, thus
one will always
assume in the sequel that $0<\alpha_{ij}<1$. In the anyon case
with bosonic (fermionic) wavefunctions, the same reasonning leads to
$0<\alpha<2$ ($-1<\alpha<1$) since then the $m_{ij}$'s are constrained
to be equal to a given even (odd)
integer.

The question is of finding particular eigenstates.
Let us confine the particles by a harmonic attraction to the origin.
This procedure [1,8] is commonly used in the anyon context since it yields
a discrete spectrum.
The $N$-particle Hamiltonian with a
harmonic interaction reads
\be \label{11}  H_N= {1\over 2m}\sum_{i=1}^{N}\left [(\vec{p}_i -
\sum_{j\ne i}\alpha _{ij}
{\vec k\times \vec r_{ij}\over r_{ij}^2})^2 + {m^2\omega^2
 {\vec r_i}^2}\right] \ee
One finds the relative eigenstates
\be \label{12} <\vec r_i|n,m_{ij}>={\cal {N}}e^{i\sum_{i<j} m_{ij}\theta_{ij}}
e^{-\beta r^2/2}
\prod _{i< j}
r_{ij}^{\vert m_{ij}-\alpha_{ij}\vert}L_{n}^{N-2+\sum_{i< j}\vert
m_{ij}-\alpha_{ij}\vert}(\beta r^2)\ee
(${\cal {N}}$ is a normalization factor,
$\beta\equiv{\displaystyle m\omega\over \displaystyle N}$ and
$r^2\equiv \sum_{i<j}r_{ij}^2$) with eigenvalues
\be\label{13} (2n+N-1+\sum_{i< j}\vert m_{ij}-\alpha_{ij}\vert)\omega\ee
Note that since the states (12) are eigenstates of the total angular momentum
operator, they are still eigenstates in the presence of a uniform magnetic
field $(\omega\to \sqrt{\omega^2+\omega_c^2})$.
The integers $m_{ij}$ have
to satisfy simultaneously either
$m_{ij}>0$ (case I) or
$m_{ij}\le 0$ (case II)\footnote {If the $\alpha_{ij}$'s are not
constrained to be in the interval $[0,1]$, one gets either
$m_{ij}>E[\alpha_{ij}]$
or $m_{ij}\le E[\alpha_{ij}]$}.
For a given $N$, the relative states $ <\vec r_i|n,m_{ij}>$
have obviously
too many quantum numbers. The $m_{1j}$'s can be
choosen as independant orbital quantum numbers, the other being either $1$
(case I) or $0$
(case II). One gets $<n,m_{ij} |n',m'_{ij}>=\delta_{n,n'}\delta_{m,m'}$
where $m=\sum_{i<j}m_{ij}$, and $ m'=\sum_{i<j}m'_{ij}$,
leading to sectors labelled by the quatum numbers $n,m$. In a given sector,
the states can be separately orthonormalized.

The
states (\ref{12}) with linear dependance on the $\alpha_{ij}$'s narrow down
to the usual linear anyonic eigenstates
when one sets $\alpha_{ij}=\alpha$. In this particular case
one has additionnal conditions on the $m_{ij}$'s depending on the statistics
(Bose or Fermi) imposed when $\alpha=0$.

One way to reproduce (\ref{12}) is to work [9] in the singular gauge
(\ref{9})
\be\label{14} \psi'=\exp(-i\sum_{k<l} \alpha_{kl}\theta_{kl}) \psi \ee
In complex coordinates $z_i=x_i+iy_i$ the free gauge transformed
Hamiltonian is
\be\label{15} H_N^{'}=\sum_{i=1}^N \left(
                          -{2\over m}\partial_{z_i} \partial_{\bar z_i}
                          +{1\over 2}m\omega^2 z_i \bar z_i\right) \ee
The states
\be\label{16} \psi'
                =\exp(-{\displaystyle N\beta\over \displaystyle 2}
                             \sum_i z_i\bar z_i) \phi \ee
are eigenstates of the Hamiltonian
(\ref{15}) if $\phi$ is an homogeneous meromorphic function
 of degree $d$ of $z_1,\dots,z_N$ (case I) or of
$\bar z_1,\dots,\bar z_N$ (case II),
with for eigenvalues $(N+d)\omega$.
Since the prefactor $\exp(-{\displaystyle N\beta\over \displaystyle 2}
\sum_{i}z_i\bar z_i)$ can be rewritten as
$\exp(-{\displaystyle \beta\over \displaystyle 2}\sum_{i<j}z_{ij}\bar z_{ij})
\exp(-{\displaystyle \beta\over \displaystyle 2}N^2 Z\bar Z)$, the center of
mass coordinate $Z=\sum_{i} z_i /N$ factorizes out
in (\ref{16}) with energy $\omega$.

A suitable basis for $\phi$ is
$\{ \prod_{i<j} z_{ij}^{m_{ij}-\alpha_{ij}} \}$ or
$\{ \prod_{i<j} \bar z_{ij}^{\alpha_{ij}-m_{ij}} \}$,
where the $m_{ij}$'s are integers, thus $ d=\sum_{i<j}(m_{ij}-\alpha_{ij})$ or
$ d=\sum_{i<j}(\alpha_{ij}-m_{ij})$ (the $m_{ij}$'s are easily seen
to be not independent
since $z_{ij}^{m_{ij}}=(z_{1j}-z_{1i})^{m_{ij}}$ can be expanded
in a power series of $z_{1i}$). The space
of eigenstates has to be a Hilbert space of square integrable functions
where the Hamiltonian is self-adjoint. A simple requirement is to
impose that the eigenstates have no divergence, implying that
one has simultaneously $m_{ij}>0$ (class I),
or $m_{ij}\le 0$ (class II).
Thus one reproduces the eigenstates
(\ref{12}) with $n=0$. An explicit calculation shows that non
vanishing radial quantum number  $n>0$ correspond to
the Laguerre polynomials appearing in (\ref{12}). The energy  $(2n+N-1
+\sum_{i<j}(m_{ij}-\alpha_{ij}))\omega$ (class I) or
$(2n+N-1
+\sum_{i<j}(\alpha_{ij}-m_{ij}))\omega$ (class II) coincides with (\ref{13}).

Imposing that the eigenstates vanish when any two particles come close together
is an anyon-like requirment, amounting to
the exclusion of the
diagonal of the configuration space.
Self-adjoint extensions [10] corresponding to diverging behavior at small
distance are possible. Indeed, allowing for short distance singularities
implies
that the $m_{ij}$'s have to satisfy simultaneously
$m_{ij}\ge 0$ or $m_{ij}\le 1$. These states diverge at the origin but are
still square integrable.
Asking for the Hamiltonian to be self adjoint
implies  additionnal constraints  leading to the
possible self adjoint extansions $m_{ij}\ge 0$ or $m_{ij}\le-1$, and
$m_{ij}\ge 2$ or $m_{ij}\le 1$.

It is amusing to note that, in the situation where the
$\alpha_{ij}$'s are either
equal to $\alpha$ or nul,
the eigenstates (\ref{12}) can be directly deduced from the
anyonic eigenstates
\be\label{17} e^{i\sum_{i<j} m_{ij}\theta_{ij}} e^{-\beta r^2/2}
\prod _{i< j}
r_{ij}^{\vert m_{ij}-\alpha\vert}L_{n}^{N-2+\sum_{i< j}\vert
m_{ij}-\alpha\vert}(\beta r^2)\ee
These states are monovalued eigenstates of the N-anyon Hamiltonian if the
integers $m_{ij}$ simultaneously satisfy $m_{ij}>E[\alpha]$ or
$m_{ij}\le E[\alpha]$ (here one cannot restrict $\alpha$ to
be in the interval [0,2]). However if one drops the monovaluedness
criterium one finds that these states are still solutions of the eigenvalue
equation if some of the
$m_{ij}$ are replaced by $m_{ij}=m'_{ij}+
\alpha$ where the integers $m'_{ij}$ have to be simultaneously $>0$
or $\le 0$. But one can get
 rid of the multivaluedness of the wavefunction by
means of the singular gauge transformation $\exp(i\alpha\sum_{i<j}\theta_{ij})$
where
the last summation is performed only on those indices $i,j$ for which $m_{ij}=
m'_{ij}+\alpha$. One then reproduces
 the eigenstates defined above where the indices $i,j$ for which
$\alpha_{ij}$ has been set to $0$ correspond to $m_{ij}=
m'_{ij}+\alpha$.


ii) Perturbative results

Leaving aside the exact eigenstates (\ref{12}), very little is known about the
model defined above. In the case where the $\alpha_{ij}$'s
are assumed to be small, a  perturbative analysis can
give some information
on the system. Here again the experience gained in the study of
the anyon model is helpful. A na\"\i ve perturbative analysis
might make no sense due to the very singular\footnote{here we
assume that the Hilbert space of
unperturbed wavefunctions does not contain any states with singular
short distance behavior.} Aharonov-Bohm interaction ${\alpha_{ij}^2/
r_{ij}^2}$.
One can circumvent this difficulty by noticing that the singular
gauge parameter $\Omega''=\sum_{i<j}\alpha_{ij}\theta_{ij}$ is the imaginary
part
of the meromorphic function
\be\label{18} \Omega(z_1,z_2,\ldots,z_N)=\sum_{i<j}{\alpha_{ij}}
\ln z_{ij} \ee
 Let us "gauge transform" the Hamiltonian $H_N$ by taking as
gauge parameter the real part of $\Omega$, $\Omega'=\sum_{i<j}\alpha_{ij}
\ln r_{ij}$
i.e.
\be\label{19} \psi
=   \exp(\pm\Omega') {\tilde{\psi}}
=\prod_{i<j} r_{ij}^{\pm \alpha_{ij}
} {\tilde {\psi}}(\vec r_1,\cdots,\vec r_N)\ee
Because of  the Cauchy-Riemann relations in  2 dimensions,
$\vec {\nabla}_i\Omega''=\vec k\times\vec {\nabla}_i\Omega'$
(implying
$\partial_{z_k}\Omega=i \bar{{\cal A}}_{k}$), one finds
that the singular terms are absent in the Hamiltonian ${\tilde {
H}_N} $ acting on ${\tilde {\psi}}$
\be\label{20}  \tilde {H}_N=\sum _{i=1}^{N}({\vec{p}_i ^2\over
2m}+\sum_{j\ne i}{i\alpha_{ij}\over m}
{\vec k \times \vec r_{ij}\over r^2_{ij}}\vec \partial_i
\mp\sum_{j\ne i}{\alpha_{ij}\over m}{\vec r_{ij}
\over r^2_{ij}}\vec \partial_i). \ee

As in the anyon model, one gets 2-body interactions with short
distance behavior adapted to a perturbative analysis.
One notes the  $\pm$ sign freedom in the choice of
the redefinition of $\tilde{\psi}$. This sign freedom describes
two possible short distance behaviors of the exact eigenstates
as emphasized in the context of self adjoint extensions. The $-$ sign
in the redefinition (\ref{19}) corresponds to the self adjoint extansion
$m_{ij}\ge 0$ (class I) or
$m_{ij}\le -1$ (class II).

iii) Some of the particles are fixed.

So far the $N$ particles are dynamical. It is however interesting
to consider some of the particles fixed in the plane.
To do
so, take for the  free Hamiltonian of
the N-particle system in the singular
gauge
\be\label{21} H^{'}_N= {1\over 2m}\sum_{i=1}^N\epsilon_i {\vec {p_i}}^2\ee
where $\epsilon_i=1$ (moving particle) or $\epsilon_i=0$ (fixed particle).
In the Lagrange formulation (\ref{1}), this amounts to affecting the speed
$\vec v_i$ with the factor $\epsilon_i$. Going back to the regular gauge
via the gauge transformation (\ref{14})
yields the desired Hamiltonian describing fixed and moving particles
interacting via the Aharonov-Bohm couplings $\alpha_{ij}$
\be\label{22} H_N=  {1\over 2m}\sum _{i=1}^{N}\epsilon_i(\vec{p}_i -
{\cal \vec A}_i)^2\ee

Most of the results presented above are still operative,
in particular the perturbative
considerations. However, the eigenstates (\ref{12})
cannot be used anymore.

Let us consider the scattering of a single
particle by a finite lattice of $N-1$ identical flux tubes. This amounts to
take
 $\epsilon_1=1$ and $\epsilon_i=0$ for $ i=2\ldots,N$.
Only the scattering by
a single flux line is solvable [2]. In the case of two flux lines
one has in the singular gauge
\be\label{23} \psi'=\exp(-i\alpha_{12}\theta_{12}-i\alpha_{13}\theta_{13})
               \psi    \ee

Let us assume that $2$ and $3$ are located
at $z_2=-h$ and $z_3=h$. For simplicity we set
$\alpha_{12}=\alpha_{13}=\alpha$.
In the limit where
the two flux lines are at the same point $h\to 0$,
one should reproduce the scattering of one electron by a flux line,
where polar coordinates are used to separate the eigenstates equation.
There exists a single coordinate system
where the eigenstate's equation separates, and which contains the polar
coordinates as a particular limiting case. This is
the elliptic coordinates system [12].
Still working in the singular gauge but with
the conformal mapping $z_1=h\cosh(\mu+i\phi)$ where
$(\mu,\phi)$ are the elliptic coordinates,
one gets the free Hamiltonian
\be\label{24}  H'= -{1\over 2 m}{1\over h^2(\cosh^2 \mu-\cos^2
\phi)}(\partial_{\mu}^2+
\partial_{\phi}^2)\ee
Notice that
the gauge transformation (\ref{23}) does not separate when one uses
elliptic coordinates. Indeed one gets
\be\label{25} \exp(i\theta_{12}+i\theta_{13})={\sinh^2(\mu+i\phi)\over
\cosh^2\mu-\cos^2
\phi}\ee

This set of coordinates has been used in [11], however we stress that
it does not
describe the scattering of a charged particle by two isolated fixed flux lines,
but instead the scattering of a charged particle
by an elliptic flux tube.
Indeed the eigenvalue equation is now separable and one can
factorize the eigenstates as $M(\mu)\Phi(\phi)$, leading
to Mathieu's equations\footnote{note that if a central harmonic
interaction is added, Hill's
equations have to be considered.} [12].
In the singular gauge,
the angular function $\Phi(\phi)$ is multivalued.
With this choice of
coordinates, the contour of the flux tube (which controls the multivaluedness
of the singular free wavefunction) must necesseraly
coincide with a geodesic defined by a constant $\mu$.
This is an ellips or in the singular case the
line $[-h,h]$ that connects $z_2=-h$ to $z_3=+h$.
It follows that the singular gauge transformation implied by the choice
of elliptic coordinates
is
$\psi'=\exp(-i\alpha\phi)\psi$.

Let us consider the singular flux $[-h,h]$ line case.
Mathieu's equations read
\be\label{26} \partial_{\phi}^2\Phi -h^2 mE\cos(2\phi)\Phi +(\lambda-h^2
mE)\Phi =0\ee
\be\label{27} \partial_{\mu}^2M +h^2 mE\cosh(2\mu)  M -(\lambda-h^2 mE)M =0\ee
where $\lambda$ is a constant introduced to separate the coordinates.
The general solution [12] for a multivalued $\Phi$ is
\be\label{28} \Phi=Ae^{{u_{\phi}}\phi}\sum_{r=-\infty}^{+\infty}
c_{2r}e^{2ir\phi}\ee
where $iu_{\phi}$ is not an integer. A possible
$\exp(-u_{\phi}\phi)\sum_{r=-\infty}^{+\infty} c_{2r}e^{-2ir\phi}$
solution has been omitted since parity is
broken anyway by the singular flux $[-h,h]$ line.  In the regular gauge, the
wavefunction has to be monovalued when a complete winding encircling the flux
$[-h,h]$ line is performed. This implies that $(u_{\phi}+i\alpha)/i$
has to be an integer $n$.  Obviously one can
assume that $n\in [0,2[$. The quantification condition on $\lambda$ comes from
the compatibility
of the homogeneous system of linear algebraic equations determining the
$c_{2r}$'s in terms of the coefficients
$${mh^2E\over 2}{1\over (2r-iu_{\phi})^2-\lambda+mh^2}$$
Thus one has two quantum numbers $\lambda$ and $E$, as desired (remember that
$u_{\phi}$ has been fixed by the monovaluedness criterium).
The solution of the equation on $M$ introduces a parameter $u_{\mu}$
which does not yield any additionnal quantum number
since there is accordingly a compatibility condition to
satisfy\footnote{ one gets $u_{\mu}=\pm i(n-\alpha) $.}.

In the limit where the flux $[-h,h]$ line shrinks to a point, $h\to 0$,
 one should reproduce the isolated flux tube case. One has
$\mu\to\infty$, and $z_1\to re^{i\phi}$ with $r=(h/2)\exp\mu$.
The Mathieu's and compatibility equations become
\be \partial^2_{\phi}\Phi+\lambda\Phi=0\ee
\be -{1\over r}\partial_r r\partial_r M+{\displaystyle \lambda\over
\displaystyle r^2} M=2mEM\ee
\be \sin^2\pi{\alpha-n\over 2}=\sin^2\pi{\sqrt {\lambda}\over 2}\ee
leading to the usual Bessel eigenfunctions
with $\lambda=(\alpha-\ell)^2$, where  $\ell=n+2p$ is the usual
angular quantum number (remember that $n=0,1$).

4. ON CURVED SPACE

i) The formalism

Let us consider a bidimensional manifold ${\cal {M}}_2$
defined by its metric $g_{ab}(x)$, where the indices $a,b$ label the space
coordinates $x^1,x^2$.
The covariant Lagrangian is given by
\be\label{32} L_N=\sum_{i=1}^N \left(
                   {mg_{ab}(x_i)\dot x_i^a
                   \dot x_i^b\over 2} +\sum_{\alpha=1}^N e_{\alpha i}
                   (g_{ab}(x_i)A_{\alpha}^a (x_i)\dot x_i^b
                   - A_{\alpha}^o(x_i))\right)
                   + \sum_{\alpha,\beta=1}^N {\kappa_{\alpha\beta}\over 2}
                   \epsilon^{\mu\nu\rho}{\displaystyle \int}
                   A_{\alpha\mu}\partial_{\nu}A_{\beta\rho} d^2x \ee
 ($g=|\det(g_{ab})|$).
It is invariant under the gauge transformation
$\delta A_\alpha^\mu=\partial^\mu\Lambda_\alpha$. By definition $[\kappa]$ is a
symmetric matrix.

The Chern-Simons topological term having not explicit dependence on the metric
one gets, by varying the Lagrangian with respect to $A_\beta^0$,
an equation for the gauge potential identical to (\ref{3})
\be\label{33}\sum_{i=1}^N e_{\beta i}\delta^2(x-x_i) =
         \sum_{\alpha=1}^N\kappa_{\alpha\beta}\epsilon^{ab} \partial_{a}
          A_{\alpha b}\ee
Defining ${\cal A}_{ia}=\sum_{\alpha} e_{\alpha i} A_{\alpha a}$
the covariant Hamiltonian reads
\be\label{34}  H_N= {1\over 2m}\sum _{i=1}^{N} {1\over\sqrt g}
                   (p_{ia} -{\cal A}_{ia}(x_i))g^{ab}{\sqrt g}
                   (p_{ib} -{\cal A}_{ib}(x_i))          \ee
where $p_{ia}={1\over i}\partial_{ia}$ are the first quantized momenta
(in the free case (\ref{34}) is the Laplace-Beltrami
operator). The gauge potentials ${\cal A}_{ia}$ are determined by
\be\label{35} \epsilon^{ab}\partial_a {\cal A}_{ib}(x)= 2\pi
                    \sum_{j}\alpha_{ij} \delta^2(x-x_j)  \ee
with $[\alpha]=[e]^t[\kappa]^{-1}[e/2\pi]$. The question is of solving
this equation on a non trivial manifold.
As on the plane one should proceed by first solving the 2-body problem
\be\label{36}
\epsilon^{ab}\partial_{ia} {\cal {A}}_{b}(x_i,x_j)
=2\pi\delta^2(x_i-x_j)  \ee
Then the vector  potentials ${\cal A}_{ia}(x_i)$
are given as a sum of 2-body terms
\be\label{37} {\cal A}_{ia}(x_i)=\sum_{j\ne i}\alpha_{ij} {\cal
A}_{a}(x_i,x_j)\ee
plus possible irrotationnal one body terms describing
the topology of the manifold (see
below the cylinder case for an illustration).
${\cal A}_a(x_i,x_j)$ is the gradient of a
potential symmetric under the exchange $x_i\to x_j$.

It follows that ${\cal A}_{ia}(x_i)$ is the gradient with
respect to $x_i$ of some function $\Omega''(x_1,\ldots,x_N)$ which is
multivalued
for any loop enclosing the singular points $x=x_j$.
In the singular gauge
\be\label{38} \psi'=\exp\left(-i\Omega''\right) \psi \ee
the Hamiltonian is free but $\psi'$
is multivalued.

In the Coulomb gauge $\partial_{a}\sqrt g {\cal A}^a=0$, one can rewrite
${\cal A}_i^a(x_i)$ as
the dual tensor $-1/{\sqrt
g}\epsilon^{ab}\partial_{ib}\Omega'(x_1,\ldots,x_N)$.
If one redefines
\be\label{39} \psi=\exp\left(\pm\Omega'\right)\tilde\psi \ee
the Hamiltonian acting on $\tilde\psi$ is
\be\label{40} \tilde H_N= \sum_{i=1}^{N}\left({1\over 2m}{1\over\sqrt g}
               p_{ia} g^{ab}{\sqrt g}p_{ib} +{\sl i}{1\over m}{\cal A}_i^a(x_i)
               \partial_{ia} \mp{1\over m}\sqrt{g} \epsilon_{ab}{\cal A}_i^a
               (x_i)\partial_i^b \right)      \ee

$\Omega'$ and $\Omega''$ are by definition harmonic for the
Laplace-Beltrami operator.
In the simple case where $g_{ab}=\pm{\sqrt g}\delta_{ab}$ (plane, cylinder
[13],
sphere [14],$\ldots$), $\Omega'$ and
$\Omega''$ satisfy the Cauchy-Riemann relation, and
$\Omega=\Omega'+i\Omega''$ is  meromorphic.

ii) An example : the cylinder

Let us consider the particular case of the cylinder
${\bf R}\times{\bf S_1}$.
One has that $x^2$ and $x^2+2\pi$
have to be identified. We define
$z=x^1+ ix^2$.
Let us first construct $\Omega$ in the 2-body case (\ref{36}). On the one hand
it must behave as $\alpha_{ij}\ln z_{ij}$ (see \ref{18}) when $z_{ij}\to 0$,
since locally the cylinder is equivalent to a plane. On the other hand
$\partial_{z_i}\Omega=i({\cal A}_{i1}-i{\cal A}_{i2})$
has to be periodic in the variable $x^2_{ij}$.
One arrives at $\Omega=\alpha_{ij}\ln(e^{z_{ij}}-1)$ which, however,
has yet to be symmetrized under
the exchange of $i$ and $j$, yielding
\be\label{41} \Omega=\alpha_{ij}\ln2\sinh{z_{ij}\over 2}\ee
One way to reproduce this result consists in considering the planar
$\Omega'=\alpha_{ij}\ln{\sqrt {(x^1_{ij})^2+(x^2_{ij})^2}}$
and making it periodic by introducing the infinite series
\be\label{42} \alpha_{ij}\sum_{n=-\infty}^\infty\ln\sqrt{(x^1_{ij})^2
                                        +(x^2_{ij}+2\pi n)^2} \ee
This series is formally divergent, but it can be given a non ambiguous meaning
by a usual procedure  (derive with respect to
$x^1_{ij}$, perform the summation, and then integrate).
One then obtains (\ref{41}).

The essential difference between the plane and the cylinder consists in
the non vanishing contour integral of a gauge field
on a non contractible loop around the cylinder.
Consequently one has to introduce a line of flux
inside the hole and, in the Coulomb gauge, consider a 1-body
multivalued term $\phi_i z_i$. In the 2-body
case this amounts to add   $\phi_i z_i+ \phi_j z_j$  to
(\ref{41})
\be\label{43} \Omega=\alpha_{ij}\ln2\sinh{{z_i-z_j}\over 2}
                       +\phi_iz_i+\phi_j z_j       \ee
Thus, in the $N$-body  case, one has
\be\label{44} \Omega=\sum_{0\le i<j\le N}\alpha_{ij}
                      \ln 2\sinh{{z_i-z_j}
                      \over 2}  +\sum_{i =1}^N \phi_iz_i       \ee

We now specialize to the 2-anyon case $\alpha_{12}=\alpha, \phi_1=\phi_2=\phi$.
It follows that the relative motion is controlled by the
anyonic $\alpha$ term, whereas
the $\phi$ term  concerns the center of mass motion.

Let us
define the center of mass $Z=(z_1+z_2)/2$ and
relative $z=z_1-z_2$  coordinates. In the singular gauge (\ref{38})
the Hamiltonian is free.
Since the configuration space of two identical particles is defined by
the identification $z\to -z$, the conformal mapping
$w=2\sinh(z/2)$ maps the cylinder on the plane and is thus well adapted. In
polar coordinates $w=r e^{i\theta}$,
the relative Hamiltonian reads
\be\label{45} H_2= -{1\over m}{\sqrt{(1-{\displaystyle 1\over \displaystyle 4}
r^2)^2
                                       +r^2\cos^2 \theta }}
                     \left({1\over r}\partial_{r}r
                     \partial_r+{1\over{r^2}}\partial_{\theta}^2
                          \right)  \ee
Going back to the regular gauge amounts
to the shift $\partial_{\theta}\to\partial_{\theta}-i\alpha$.

Thus one has a situation similar
to the relative scattering of two anyons on a
plane. However, the Jacobian of the
conformal mapping makes the relative motion
non separable, implying seemingly out of reach exact eigenstates.


\bf {Acknowledgments :} \rm
S.O. acknowledges useful discussions with A. Comtet and
J. Stern.  The two of us acknowledge stimulating conversations with J. McCabe.

\vfill\eject

\bf {References}\rm

[1] J.M. Leinaas and
    J. Myrheim,  Nuovo  Cimento B {\bf 37}, 1  (1977);  J.M.  Leinaas,  Nuovo
    Cimento A {\bf 47}, 1 (1978); M.G.G. Laidlaw and  C.M. de Witt, Phys.
    Rev. D{\bf 3}, 1375 (1971)

[2] Y. Aharonov and D. Bohm, Phys. Rev. {\bf 115} (1959) 485;
in the case of two flux tubes see for example
    P. \v S\v tov\'\i\v cek, Phys. Lett. {\bf A} 142, 1 (1989) 5

[3] F.  Wilczek, Phys.  Rev.  Lett. {\bf 49}, 957 (1982);
    Y.S.Wu, Proc. $2^{nd}$ Int. Symp. Foundations of
    Quantum Mechanics, Tokyo,  171 (1986) and
    Phys. Rev. Lett.  {\bf 53} (1984); for a general review on the subject
    see F. Wilczek,  "Fractional Statistics and Anyon Superconductivity",
    World Pub. (1990); Y.-H. Chen, F. Wilczek, E. Witten, et
    B. I. Halperin, Int. J. Mod. Phys. B 3 (1989) 1001

[4]  Y. S. Wu, Phys. Rev. Lett. {\bf 53}, 111 (1984); G.V. Dunne, A. Lerda and
C.A. Trugenberger, Mod. Phys. Lett. A {\bf 6}, 2891 (1991);
     ibid, Int. Jour. Mod. Phys. B {\bf 5},  1675 (1991);   see also G.V.
Dunne, A. Lerda , S. Sciuto and C.A. Trugenberger, ``Exact multi-anyon
     wavefunctions in a magnetic field '', preprint CTP\#1978-91;
     J. Grundberg, T.H. Hansson, A. Karlhede and E. Westerberg,
     ``Landau levels for anyons'', preprint USITP-91-2;
     A.P. Polychronakos, Phys. Lett. B {\bf 264}, 362 (1991);
     C. Chou, Phys. Lett. A {\bf 155}, 245 (1991); K. H. Cho and C. Rim,
     `` Many anyon wavefunctions in a
     constant magnetic field'', preprint SNUTP-91-21;
     A. Khare, J. McCabe and S. Ouvry, Phys. Rev. D{\bf46   } (1992) 2714;
     G.V. Dunne, A. Lerda , S. Sciuto and C.A. Trugenberger, `` Magnetic
     Moment and Third
     virial of non-relativistic anyons '', preprint MIT-CTP 2032, CERN-TH 6338;
     E. Verlinde, `` A Note on Braid Statistics and the Non-Abelian
     Aharonov-Bohm Effect'', preprint IASSNS-HEP-90/60

[5] J. McCabe and S. Ouvry , Phys. Lett. B {\bf 260} (1991) 113;    A. Comtet,
  J. McCabe and S. Ouvry , Phys. Lett. B {\bf  260} (1991) 372;
   A. Dasni\a`eres de Veigy and S. Ouvry, Phys. Lett. {\bf 291B} (1992) 130;
 Nucl. Phys. {\bf B[FS] 388} (1992) 715; see also
   D. Sen, Nucl. Phys. B {\bf 360 } (1991) 397;
   C. Chou, L. Hua and G. Amelino-Camelia, Phys. Lett. B {\bf 286} (1992) in
press;
G. Amelino-Camelia, Phys. Lett. B {\bf 286} (1992) 97

[6] M.V.N. Murthy, J. Law, M. Brack and R.K. Bhaduri, Phys. Rev. Lett. {\bf
67},
 1817 (1991);
M. Sporre, J.J.M. Verbaarschot and I. Zahed,  Phys. Rev. Lett. {\bf 67}, 1813
(1991); ibid, `` Four anyons in an
harmonic well '', preprint SUNY-NTG-91/40; `` Anyon Spectra and the Third
Virial
Coefficient'', preprint SUNY-NTG-91/47; J. Law, M. Suzuki and R.K. Bhaduri,
Phys. Rev. {\bf A46} (1992) 4693; J. Myrheim and K. Olaussen, "The Third Virial
Coefficient of Free Anyons" Unit Trondheim report (1992)

[7] J. Fr\"ohlich and A. Zee, Nucl. Phys. {\bf B[FS] 364} (1991) 517;
    X. G. Wen, A. Zee, Phys. Rev. {\bf B 46}, 4 (1992) 2290

[8] A. Comtet, Y. Georgelin and S. Ouvry, J. Phys. A : Math. Gen.
    {\bf 22}, 3917-3925 (1989)

[9]  See for example G. V. Dunne et al in [2]

[10]  R. Jackiw, MIT preprint CTP1937 (1991); C. Manuel and R. Tarrach, Phys.
Lett. B{\bf 268} (1991) 22; J. Grundberg, T.H. Hansson, A. Karlhede and J. M.
Leinaas, Mod. Phys. Lett. B {\bf 5} (1991) 539

[11]  Z. Y. Gu and S. W. Qian, J. Phys. {\bf A} : Math. Gen.
                                              {\bf 21} (1988) 2573

[12] I. S. Gradshteyn and I. M. Ryzhik, "Table of integrals,
     series and products", Academic Press, New York and London (1965) p.
     991

[13] for alternative studies of the anyon model on a cylinder see
Y. S. Wu, Int. J. Mod. Phys. {\bf B 5}, 10 (1991) 1649;
     Y. Hatsugai, M. Kohmoto and Y. S. Wu, Phys. Rev. B {\bf 43} (1991) 2661;
S. Chakravarty and Y. Hosotani, Phys. Rev. D{\bf 44} (1991) 441

[14] A. Comtet,
  J. McCabe and S. Ouvry, Phys. Rev. D{\bf 45}  (1992) 709;
D. Li, "Anyons and Quantum Hall Effect on the Sphere" Sissa Report 129/91/EP



\end{document}